\documentclass[12pt]{article}

\usepackage{amsmath,amsfonts,amssymb,amsbsy,textcomp}
\usepackage{amsmath,euscript,array,cite,mathrsfs}
\usepackage{graphicx,color}
\usepackage{hyperref}
\usepackage{graphics}
\usepackage{epsfig}
\usepackage{verbatim}
\usepackage{amsfonts}

\usepackage{epsf}

\def\be{\begin{equation}}
\def\ee{\end{equation}}
\def\bea{\begin{eqnarray}}
\def\eea{\end{eqnarray}}
\def\bc{\begin{center}}
\def\ec{\end{center}}

\def\cC{{\mathcal{C}}}

\def\cH{{\mathcal{H}}}

\def\cN{{\mathcal{N}}}

\def\cW{{\mathcal{W}}}

\def\cO{{\mathcal{O}}}

\def\ep{{\epsilon}}

\def\r2{{\sqrt{2}}}

\def\r{\rho}

\newcommand{\Tr}{\operatorname{Tr}}

\def\B0{{\boldsymbol 0}}

\def\Tr{{\rm Tr}}

\def\ee{\boldsymbol{e}}

\def\Dbarslash{\,\,{\raise.15ex\hbox{/}\mkern-12mu {\bar D}}}
\def\Dslash{\,\,{\raise.15ex\hbox{/}\mkern-12mu D}}
\def\delslash{\,\,{\raise.15ex\hbox{/}\mkern-9mu \partial}}
\def\delbarslash{\,\,{\raise.15ex\hbox{/}\mkern-9mu {\bar\partial}}}

\newcommand{\EQ}[1]{\begin{equation}\begin{split} #1
\end{split}\end{equation}}

\makeatletter \@addtoreset{equation}{section} \makeatother

\begin{document}

\begin{titlepage}

\begin{flushright}
DAMTP-2011-26\\
MAD-TH-11-01\\

\end{flushright}
\begin{centering}

\vspace{.6in}

{ \Large{\bf A New 2d/4d Duality via Integrability}} 

\vspace{.2in}

Heng-Yu Chen${}^{1}$, Nick Dorey${}^{2}$, Timothy J. Hollowood${}^{3}$ and
Sungjay Lee${}^{2}$
\\
\vspace{.3 in}
${}^{1}$Department of Physics,\\
University of Wisconsin-Madison,\\
Madison, WI 53706, USA
\\
\vspace{.1 in}
${}^{2}$DAMTP, Centre for Mathematical Sciences,\\
University of Cambridge, Wilberforce Road,\\
Cambridge, CB3 0WA, UK
\\
\vspace{.1in}
${}^{3}$Department of Physics,\\
Swansea University,\\
Swansea SA2 8PP, UK
\\
\vspace{.5in}

{\bf \large Abstract} \\
\end{centering}
We prove a duality,
recently conjectured in arXiv:1103.5726, which relates the F-terms of
supersymmetric gauge theories defined
in two and four dimensions respectively.
The proof proceeds by a saddle point analysis of the
four-dimensional partition function in the Nekrasov-Shatashvili
limit. At special quantized values of the
Coulomb branch moduli, the saddle point condition
becomes the Bethe Ansatz Equation of the $SL(2)$
Heisenberg spin chain which coincides with the F-term equation of the dual
two-dimensional theory. The on-shell values of the superpotential in the two
theories are shown to coincide in corresponding vacua.
We also identify two-dimensional duals for a large set of quiver
gauge theories in four dimensions and generalize our proof to
these cases.

\vspace{.1in}

\end{titlepage}

\renewcommand{\thefootnote}{\#\arabic{footnote}}
\setcounter{footnote}{0}

\section{Introduction}

Two dimensional theories have long been studied as toy
models for aspect of four-dimensional gauge dynamics such as
asymptotic freedom, instanton effects, the generation of a mass gap
and large-$N$ limits. Recently a duality between two- and four-dimensional
theories was conjectured \cite{DHL} which makes this analogy precise
for some protected quantities in the supersymmetric setting.
The proposed duality relates four-dimensional ${\cal N}=2$ gauge
theories in a particular
$\Omega$ background
to ${\cal N}=(2,2)$ gauged linear sigma models in two dimensions.
The new duality extends an earlier proposal \cite{Dorey1998,Dorey1999,Lee2009}
which related the BPS spectrum of ${\cal N}=(2,2)$ QED with charged
matter to that of $SU(N)$ Seiberg-Witten theory with massive flavors at the
Higgs branch root. In particular it can be regarded as an extension of
the earlier proposal away from the Higgs branch root which holds
at a generic point on the Coulomb
branch of the four-dimensional theory. The proposal also makes contact
with another, quite different type of 2d/4d duality: the AGT
conjecture \cite{AGT} which relates the instanton partition functions of
four-dimensional
${\cal N}=2$ superconformal theories to conformal blocks of
Liouville theories on Riemann surfaces.
In this letter, we will present a proof of the conjecture of
\cite{DHL} and also extend the duality
to a larger class of ${\cal N}=2$ quiver gauge theories in
four dimensions.

Let us begin by recalling the two specific theories of the aforementioned
duality, which we shall refer to as Theory I and II.

\noindent{\bf Theory I}: Four Dimensional $\cN=2$ SQCD with gauge group
$SU(L)$,
with $L$ fundamental hypermultiplets of masses $\vec m_F=(m_1,...,m_L)$
and $L$ anti-fundamental hypermulitplets of masses
$\vec m_{AF}=(\tilde m_1,...,\tilde m_L)$.
The marginal coupling constant is $\tau=4\pi
i/g^2+\vartheta/2\pi$.

\noindent Theory I is now also subjected to a particular Nekrasov deformation
on
one-plane with the deformation parameters $(\epsilon_1,\epsilon_2)=(\epsilon,
0)$,
which preserves ${\cal N}=(2,2)$ supersymmetry in a two-dimensional subspace
of four-dimensional space-time \cite{NekSha1}.
This Nekrasov deformation, or $\Omega$-background,
turns out to lift the Coulomb branch moduli space of the given theory,
leaving isolated vacua at the points,
\begin{align}\label{vacI}
  \vec a = \vec m_F - \vec n \epsilon\ ,
\end{align}
where $\vec a$ are the usual special K\"{a}hler coordinates on the
Coulomb branch and
$\vec n = (n_1,n_2,..,n_L) \in {\mathbb Z}^L$. In the presence
of the deformation, the partition function of
Nekrasov provides a twisted superpotential $\cW^{(I)}$ that describes the
low-energy dynamics of Theory I and whose critical points are given by
(\ref{vacI}).

The other system of interest is,

\noindent{\bf Theory II:} Two dimensional $\cN=(2,2)$ supersymmetric Yang-Mills
with gauge group $U(N)$,
with $L$ fundamental chiral multiplets with twisted masses $\vec M_F=
(M_1,...M_L)$ and $L$ anti-fundamental chiral multiplets with twisted masses
$\vec M_{AF}=(\tilde M_1, ...,\tilde M_L)$ as well as a single adjoint chiral
multiplet
with twisted mass $\epsilon$. The FI parameter $r$ and 2d vacuum angle $\theta$
also
combine to give a holomorphic coupling constant
$\hat{\tau}=ir +\theta/2\pi$.

\noindent As explained in \cite{DHL},
Theory II arises as the worldvolume theory of surface
operators/vortex strings
which probe the Higgs branch of Theory I. The low-energy dynamics of Theory II
is
also characterized by a twisted superpotential $\cW^{(II)}$ whose vacuum
conditions
takes the following form
\EQ{
\prod_{l=1}^L\frac{\lambda_j-M_l}{\lambda_j-\tilde M_l}= -q
\prod_{k=1}^N\frac{\lambda_j-\lambda_k-\epsilon}{\lambda_j-\lambda_k+\epsilon}\
,
\qquad q = (-1)^{N+1} e^{2\pi i \hat \tau}\ .
\label{vacII}
}
Here $\{\lambda_j\}$ represent vacuum expectation values of
the scalar field in the vector multiplet,
while the condition (\ref{vacII}) coincides with the Bethe Ansatz
Equations (BAEs) of the $SL(2)$ Heisenberg spin
chain, $\{\lambda_i\}$ being associated with magnon rapidities or
``Bethe roots''.
The vacuum equation allows
non-degenerate vacua, parameterized again by a set of integers $\hat n_l$ with
$N=\sum_{l=1}^L \hat n_l$, whose weak-coupling expressions become
\EQ{
\lambda_{(ls)}=M_l-(s-1)\epsilon+{\cal O}(q)\ , \qquad
s=1,...,\hat n_l\ .
\label{bethestring}
}
According to \cite{CV1}, massive theories preserving two-dimensional
${\cal N}=(2,2)$ supersymmetry can be classified by their critical values of
(twisted) superpotentials\footnote{A massive theory is defined
to
have a mass gap with non-degenerate vacua. Authors have also discussed
a refined classification of two-dimensional theories by their degeneracies of BPS spectra.}.
In \cite{DHL}, it has been checked, up to first few orders of instanton
expansion in $q=e^{2\pi i\tau}$
that there is an one-to-one correspondence between the supersymmetric vacua of
two theories.
Moreover it has been further conjectured that the on-shell values of their
twisted superpotentials coincide:
\EQ{
 \mathcal{W}^{\text{(I)}}(a_l=m_l-n_l\epsilon) -{\cal
W}^\text{(I)}(a_l=m_l-\epsilon)
 \equiv
 \mathcal{W}^ {\text{(II)}}(\{\hat n_l \})\ ,
 \label{ytt}
 }
provided the parameters in both theories are identified as follows
\footnote{The second term in (\ref{ytt}) is a vacuum
independent subtraction which ensures that the superpotential vanishes
at the Higgs branch root.}:
\begin{align}\label{Id1}
  \hat \tau = \tau + \frac12 (N+1)\ , \qquad
  \vec M_F = \vec m_F - \frac32 \epsilon\ , \qquad
  \vec M_{AF} = \vec m_{AF} + \frac12 \epsilon
\end{align}
with $\hat n_l = n_l -1$. In other words, protected holomorphic structures of
Theory I and Theory II are isomorphic. In particular, two theories have
the same chiral ring structure. The explicit identifications between
the chiral rings of the two theories will be discussed further below.

Four-dimensional ${\cal N}=2$ supersymmetric gauge theories
in the $\Omega$-background with $(\epsilon_1,\epsilon_2)=(\epsilon,0)$
have been studied by Nekrasov and Shatashvili \cite{NekSha1}
in relation to the quantum integrable systems. In particular,
the generators of the twisted chiral ring are mapped to quantum Hamiltonians.
It is known that the Seiberg-Witten curve of Theory I is nothing but
the spectral curve of the classical
$SL(2,{\mathbb R})$ spin chain \cite{SW2, Hanany1995}.
As above, the vacuum equations (\ref{vacII}) of Theory II can be
identified
as the BAEs of the same spin chain, where
the parameter $\epsilon$ plays a role as the Planck constant $\hbar$.
The duality therefore supports the idea of Nekrasov and Shatashvili
and may shed new light on the quantisation of integrable systems.

In order to prove this duality (\ref{ytt}), we rely on the saddle point
analysis of the Nekrasov partition function of Theory I in the $\epsilon_2\to 0$ limit,
developed recently in \cite{Maruyoshi:2010iu,Italians1, Pog}. More precisely,
we will see how the Bethe Ansatz Equation (BAE) of $SL(2,{\mathbb R})$ spin
chain
can arise from the saddle point equations of the instanton partition function.
As a consequence we can also show that the on-shell Nekrasov partition function
, ${\cal W}^\text{(I)}$ agrees with the on-shell Yang-Yang potential \cite{YY}
of
$SL(2,{\mathbb R})$ spin chain, ${\cal W}^{\text{(II)}}$.
Applying the same analysis, we can prove the duality for a
large class of linear quiver gauge theories.

\section{BAE from Nekrasov Instanton Partition Function}

There are several ways to present Nekrasov's extraordinary result for the
instanton partition function of an ${\cal N}=2$ gauge theory. Our starting
point will be the gamma function representation for the instanton partiton function
in the ${\cal N}=2$ gauge theory with $2L$ fundamental hypermultiplets
\cite{Nekrasov:2003rj}. The expression depends on a sum over $L$ Young Tableaux
$\vec Y=(Y_1,\ldots,Y_L)$. The number of boxes in $i^\text{th}$ row of the
tableau $Y_l$ ($l=1,2,..,L$) is denoted $k_{li}$ and $|\vec Y|$ is the total number of boxes
in all the $L$ tableaux. In the following,
$q=e^{2\pi i\tau}$ is the coupling  and we have defined (following
\cite{Italians1})
\EQ{
x_{li}=a_l+(i-1)\epsilon_1+\epsilon_2k_{li}\ ,\qquad
x_{li}^{(0)}=a_l+(i-1)\epsilon_1\ ,
\label{xdef}
}
where $i,j$, {\it etc.\/}, are indices that range from 1 to $\infty$.
The partition function involves a sum over the $L$ tableaux,
\EQ{
{\cal Z}_\text{inst}=\sum_{\vec Y}q^{|\vec Y|}
{\cal Z}_\text{vec}(\vec Y)\prod_{n=1}^{2L}{\cal Z}_\text{hyp}(\vec Y,\mu_n)\ ,
\label{aww}
}
where the contribution from the vector multiplet can be written
\EQ{
{\cal Z}_\text{vec}(\vec Y)=
\prod_{(li)\neq(nj)}\frac{
\Gamma\big(\epsilon_2^{-1}(x_{li}-x_{nj}-\epsilon_1)\big)}{
\Gamma\big(\epsilon_2^{-1}(x_{li}-x_{nj})\big)}\cdot
\frac{\Gamma\big(\epsilon_2^{-1}(x^{(0)}_{li}-x^{(0)}_{nj})\big)}
{\Gamma\big(\epsilon_2^{-1}(x^{(0)}_{li}-x^{(0)}_{nj}-\epsilon_1)\big)}
\label{vmm}
}
and the contribution from a single fundamental hypermultiplet of mass $\mu$ is
\EQ{
{\cal Z}_\text{hyp}(\vec
Y,\mu)=\prod_{li}\frac{\Gamma\big(\epsilon_2^{-1}(x_{li}+\mu)\big)}{\Gamma\big(\epsilon_2^{-1}(x_{li}^{(0)}+\mu)\big)}\
.
}
In order to agree with the conventions of \cite{DHL}, we take our $2L$
hypermultiplets to have masses $\{-m_l+\epsilon_1,-\tilde m_l\}$.

Now we consider the Nekrasov-Shatashvili limit $\epsilon_2\to0$ with
$\epsilon\equiv\epsilon_1$ fixed. In this limit, we can approximate the gamma
functions using Stirling's approximation, to find the leading order behaviour
\EQ{
{\cal Z}_\text{vec}(\vec Y)&=
\exp\Big[\frac1{2\epsilon_2}\sum_{(li)\neq(nj)}
\Big(f(x_{li}-x_{nj}-\epsilon)-f(x_{li}-x_{nj}+\epsilon)\\ &\qquad~~~~~~~~~-
f(x_{li}^{(0)}-x_{nj}^{(0)}-\epsilon)+f(x_{li}^{(0)}-x_{nj}^{(0)}+\epsilon)\Big)\Big]
}
and
\EQ{
{\cal Z}_\text{hyp}(\vec Y,\mu)=
\exp\Big[\frac1{\epsilon_2}\sum_{li,n}
\Big(f(x_{li}+\mu)-f(x_{li}^{(0)}+\mu)\Big]\ ,
}
where $f(x)=x(\log x-1)$.
The coupling constant piece can then be written as
\EQ{
q^{|\vec Y|}=\exp\Big[\frac{\log
q}{\epsilon_2}\sum_{li}\big(x_{li}-x_{li}^{(0)}\big)\Big]\ .
}
In the NS limit, $\epsilon_2k_{li}$ becomes continuous and so the sum over
Young Tableaux can be traded for an integral over the infinite set of variables
$\{x_{li}\}$ and we can write
\EQ{
{\cal Z}_\text{inst}=\int \prod_{li}dx_{li}\,\exp\Big[\frac1{\epsilon_2}
{\cal H}_\text{inst}(x_{li})\Big]\ ,
\label{pfu}
}
where the instanton action functional takes the difference form
\EQ{
{\cal H}_\text{inst}(x_{li})={\cal Y}\big(x_{li}\big)-{\cal
Y}\big(x_{li}^{(0)}\big)\ ,
\label{loo}
}
where
\EQ{
{\cal Y}\big(x_{li}\big)&=\log q\sum_{li}x_{li}+\sum_{li,n}\big(f(x_{li}-\tilde
{m}_n)+f(x_{li}-m_n+\epsilon)\big)\\
&\qquad\qquad+\frac12\sum_{(li)\neq(nj)}\big(f(x_{li}-x_{nj}-\epsilon)-f(x_{li}-x_{nj}+\epsilon)\big)
\ .\label{DefWx}
}

In order to make contact with \cite{NekSha1,Italians1}, we can write the
instanton action functional in integral form by introducing the instanton
``density" $\rho(x)$ which is constant along the
series of  intervals
\EQ{\label{interval}
{\EuScript I}=\bigcup_{li}\,[x_{li}^{(0)},x_{li}]\ .
}
More precisely, these are contours in the complex plane with end points
$x_{li}$ and $x_{li}^{(0)}$. Then using the identity
\EQ{
\sum_{i=1}^\infty\big(f(y-x_{li}^{(0)}-\epsilon)-f(y-x_{li}^{(0)}+\epsilon)\big)
=f(y-a_l-\epsilon)+f(y-a_l)\ ,
}
one can show
\EQ{\label{Hinst}
{\cal H}_{\rm inst}[\rho]=-\frac12\int dx\,dy\,\rho(x)\mathfrak G(x-y)\rho(y)
+\int dx\,\rho(x)\log\big(q\,{\mathfrak R}(x)\big)\ .
}
Here, the integration kernel is given by
\EQ{\label{Gx}
{\mathfrak G}(x)=\frac{d}{dx}\log\Big(\frac{x-\epsilon}{x+\epsilon}\Big)
}
and
\EQ{
\mathfrak R(x)=\frac{A(x)D(x+\epsilon)}{P(x)P(x+\epsilon)}\ ,
}
with
\EQ{
A(x)=\prod_{l=1}^L(x-\tilde{m}_l)\ ,\qquad D(x)=\prod_{l=1}^L(x-m_l)\ ,\qquad
P(x)=\prod_{l=1}^L(x-a_l)\ .
}

In the Nekrasov-Shatashvili limit
$\epsilon_2\to0$,  the functional integral (\ref{pfu}) is dominated by a saddle
point configuration; variation of the instanton density $\rho(x)$ can be
effectively achieved by small variation of end points $x_{li}$ of ${\EuScript I}$
where $\rho(x)$ should remain constant. The saddle point equation then becomes
\EQ{
\frac{\delta \cH_{\rm inst}[\rho]}{\delta x_{li}}
=-\int_{\mathfrak I} dy\,\mathfrak G(x_{li}-y)\rho(y)+\log\big(q\,\mathfrak
R(x_{li})\big)=0\ .
}
Since $\mathfrak G(x)$ is a total derivative, we can easily rewrite
the above equation into a following form
\EQ{
\frac{\mathfrak Q(x_{li} +\epsilon)\mathfrak Q^{(0)}(x_{li}-\epsilon)}
{\mathfrak Q(x_{li}-\epsilon)\mathfrak Q^{(0)}(x_{li}+\epsilon)}=-q\, \mathfrak
R(x_{li})
}
where
\EQ{
\mathfrak Q(x)=\prod_{l=1}^L \prod_{i=1}^\infty (x-x_{li})\ ,\qquad\mathfrak
Q^{(0)}(x)=\prod_{l=1}^L \prod_{i=1}^\infty  (x-x_{li}^{(0)})\ .
}
Using the explicit expression for $x_{li}^{(0)}$ in (\ref{xdef}), one can further
simplify the saddle point equation as follows
\begin{align}
  \frac{\mathfrak Q(x_{li}+\epsilon)}
  {\mathfrak Q(x_{li}-\epsilon)}=-q\, A(x_{li})D(x_{li}+\epsilon)\ .
  \label{lss}
\end{align}

The above equations \eqref{lss} are an infinite set of equations for the end-points
$x_{li}$ of internals ${\EuScript I}$. Notice that these equation do not depend on
$x_{li}^{(0)}$; however, they must be solved subject to the condition that the
solution has the expansion
\EQ{
x^{(0)}_{li}=a_l+(i-1)\epsilon_1\ ,\qquad
x_{li}=x_{li}^{(0)}+\sum_{k=i}^\infty q^kx_{li}^{(k)}\, \qquad
i=1,\dots,\infty\ ,
}
in order that the instanton partition function has a consistent expansion in
$q$. It is important that order of $q$ correlates with the index $i$, so that
at any given order in the instanton
expansion ${\cal O}(q^k)$, we can effectively truncate the infinite system of
equations by taking $x_{li}=x_{li}^{(0)}$, for $i>k$.
It is rather remarkable that the equations \eqref{lss} related to a
quantization of the Seiberg-Witten curve that is recovered in the limit
$\epsilon\to0$ \cite{Italians1, Pog}, which we will discuss later.

The important result we now want to verify is that the infinite set of
saddle-point equations has a natural truncation to a finite system if one
imposes quantization conditions on the VEVs $a_l$
\EQ{
a_l=m_l-n_l\epsilon\ , \quad n_l\in {\mathbb Z}>0\ .
\label{quant}
}
More precisely, one can show from \eqref{quant} and \eqref{lss} that
most of intervals in ${\EuScript I}$ become degenerate
\EQ{
x_{li}=x_{li}^{(0)}=a_l+(i-1)\epsilon\ ,\text{  for  }i\geq n_l\ ,
}
which leads to collapsing of the infinite set of saddle point equations
onto a finite set of equations. We will present a formal proof of this
statement because of its central role in our analysis.

\noindent {\bf Proof:}
Following \cite{Italians1}, we define
\EQ{
w(x)=\frac{\mathfrak Q(x-\epsilon)}{\mathfrak Q(x)}\ .
}
One can then rewrite the saddle point equation \eqref{lss} as
\EQ{
1+qA(x_{li})D(x_{li}+\epsilon)w(x_{li})w(x_{li}+\epsilon)=0\ .
\label{kii}
}

For later convenience, let us consider a function T(x)
\EQ{
T(x)=\frac{h+2}{w(x+\epsilon)} \Big[ 1 - \frac{h}{h+2} A(x)D(x+\epsilon)w(x) w(x+\epsilon)
\Big]\ ,
\label{dff}
}
where $q = - \frac{h}{h+2}$.
Using \eqref{kii}, we can see that apparent poles in $T(x)$ coming
from the zeros of $w(x+\epsilon)$ are cancelled by corresponding zeros
in the numerator.  It implies that $T(x)$ is analytic in the complex plane.
From the asymptotic behavior of $w \sim x^{-L}$ at large $x$ \cite{Italians1},
one can conclude that $T(x)$ should be a polynomial of degree $L$.
In the limit $\epsilon\to 0$, (\ref{dff}) reduces to a defining
equation of the Seiberg-Witten curve for ${\cal N}=2$ $SU(L)$ SQCD
with $N_{F}=2L$ fundamental flavours \cite{SW2, Hanany1995}
\EQ{
t^2- T(x) t - h ( h+2) A(x)D(x)=0\ , \qquad
t= \frac{h+2}{w(x)}\ .
\label{swc}}
In particular, the coefficients in the polynomial function
$T(x)$ correspond to the Coulomb branch moduli. It strongly
suggests that, with finite $\epsilon$, \eqref{dff} can now be interpreted as a
quantization of the Seiberg-Witten curve \cite{Italians1}.

It follows from the above that
\EQ{
{\cal A}(x+\epsilon)-{\cal B}(x){\cal A}(x)=-q\,\mathfrak R(x){\cal
A}(x-\epsilon)\ ,
\label{qws}
}
where
\EQ{
{\cal A}(x)=\frac{\mathfrak Q(x)}{\mathfrak Q^{(0)}(x)}\ ,\qquad{\cal
B}(x)=\frac{1}{(h+2)}\frac{T(x)}{P(x+\epsilon)}
}
Notice that ${\cal A}(x)$ has poles at $x_{li}^{(0)}$. Now generically both
sides of \eqref{qws}, have poles at $a_l+(i-2)\epsilon$, $i=1,2\ldots,\infty$.
But if the
quantization condition \eqref{quant} is imposed then the pole on the right-hand
side at
$a_l+(n_l-1)\epsilon$ is missing because then $\mathfrak R(x)$ has a zero
there. Consequently, on the left-hand side, ${\cal A}(x)$ cannot have a pole at
$a_l+(n_l-1)\epsilon$. But then the right-hand side does not have a pole at
$a_l+n_l\epsilon$ implying on the left-hand side ${\cal A}(x)$ cannot have a
pole at $a_l+n_l\epsilon$. The argument continues inductively for $i \ge n_l$
and the conclusion is that ${\cal A}(x)$ only has a finite set of poles
at $a_l+(i-1)\epsilon$, for $i=1,2,\ldots,n_l-1$. This implies that
\EQ{
x_{li}=x_{li}^{(0)}=a_l+(i-1)\epsilon\ ,\text{  for  }i\geq n_l\ ,
}
so only the first $n_l-1$ rows of the Young tableau $Y_l$ are occupied. This
completes the proof. \hfill $\blacksquare$

As a consequence of the above truncation, the quantised Seiberg-Witten curve
indeed can be identified as the Baxter equation of our interest. The details
of it are in order. Defining a finite polynomial
\begin{align}
  \hat{\mathfrak Q}(x)=\prod_{l=1}^L \prod_{i=1}^{n_l-1} (x-x_{li})\ ,
  \label{tru}
\end{align}
one can show that
\begin{align}
  w(x) = & \frac{\hat{\mathfrak Q}(x-\epsilon)}{\hat{\mathfrak Q}(x)} \prod_{l=1}^L
  \frac{1}{x- a_l - (n_l-1)\epsilon}
  \nonumber \\ = &
  \frac{\hat{\mathfrak Q}(x-\epsilon)}{\hat{\mathfrak Q}(x)} \cdot
  \frac{1}{D(x+\epsilon)}\ ,
\end{align}
where we used for the last equality the quantisation condition (\ref{quant}).
The quantised Seiberg-Witten curve (\ref{dff}) can then be simplified as follows
\begin{align}
  T(x) \hat{\mathfrak Q} (x) =  (h+2) D(x+2\epsilon) \hat{\mathfrak Q} (x+ \epsilon)
  - h A(x) \hat{\mathfrak Q} (x-\epsilon)\ ,
  \label{adf}
\end{align}
while the saddle point equations (\ref{kii}) become
\begin{align}
  \frac{D(x_{li}+2\epsilon)}{A(x_{li})}=-q\frac{\hat{\mathfrak Q}
  (x_{li}-\epsilon)}{\hat{\mathfrak Q}(x_{li}+\epsilon)}\ .
  \label{baa}
\end{align}
In order to make the identification of two equations (\ref{adf},\ref{baa})
with the Baxter equation and BAE of the $SL(2,\mathbb{R})$, let us
apply the identification of the mass parameters given in (\ref{Id1}) and
set $\lambda = x + \frac12 \epsilon$. It leads to
\begin{align}
  \hat{\mathfrak Q} (x) = Q(\lambda)= \prod_{l=1}^L \prod_{i=1}^{n_l-1}(\lambda - \lambda_{li})\ ,
  \qquad \lambda_{li} = x_{li} + \frac12 \epsilon\ ,
\end{align}
and
\EQ{
A(x)=a(\lambda)=\prod_{l=1}^L(\lambda-\tilde{M}_l)\,,\qquad
D(x+2\epsilon)=d(\lambda)=\prod_{l=1}^L(\lambda-M_l)\ .
\label{sqq}
}
One can finally show that (\ref{adf}) can be rewritten as a standard form of
the Baxter equation for the spin chain
\begin{align}
  t(\lambda) Q(\lambda) = ( h+2) d(\lambda) Q(\lambda + \epsilon)
  - h a(\lambda) Q(\lambda - \epsilon)\ ,
\end{align}
where $t(\lambda) = T(x)$ can be understood as the eigenvalue
of the spin chain transfer matrix. One can also see that (\ref{dff})
are precisely the BAE of $SL(2,\mathbb{R})$ spin chain
\EQ{
\frac{d(\lambda_{li})}{a(\lambda_{li})}=-q\frac{Q(\lambda_{li}-\epsilon)}
{Q(\lambda_{li}+\epsilon)}\ . \label{Spinchain1}
}
It is noteworthy here that the finite instanton string $x_{li}^{(0)}=a_l+(i-1)\epsilon$
($i=1,\ldots,n_l-1$) can be identified with the classical Bethe string solution,
\EQ{
\lambda_{(ls)}^{(0)}=M_l-(s-1)\epsilon\ ,\qquad s=1,2,\ldots,\hat n_l\ ,
}
with $\hat n_l=n_l-1$ and $s=n_l-i$.

By explicit evaluation of the instanton action with the quantization condition
and subsequent truncation, we can go one step further to show
how the Yang-Yang functional $Y(\lambda_j)$ of the spin chain (\ref{Spinchain1}),
twisted superpotential of the two-dimensional theory, can arise from the above analysis.
Denoting $N=\sum_{l=1}^L n_l -1 $, it follows from \eqref{loo}
that the instanton action in the truncated theory takes the form
\EQ{
{\cal W}^ {\text{(I)}}_\text{inst}(m_l-n_l\epsilon)=
\hat{\cal Y}\big(x_{li}\big)-\hat{\cal Y}(x_{li}^{(0)}\big)\ ,\label{onshellH}
}
where the function $\hat{\cal Y}(x)$ is a truncated version of ${\cal Y}(x)$
\EQ{
\hat{\cal Y}(x_{li})=& \log
q\sum_{(li)=1}^N x_{li}+\sum_{(li)=1}^N\sum_{n=1}^L\Big(f(x_{li}-\tilde
{m}_n)-f(x_{li}-m_n+2\epsilon)\Big)\\ &
+\frac12\sum_{(li)\neq (mj)=1}^N \Big(f(x_{li}-x_{mj}-\epsilon)-
f(x_{li}-x_{mj}+\epsilon)\Big)\ .
\label{DefWx}
}
If we make the parameter identification (\ref{Id1}) and change of variable
$\lambda=x + \frac12\epsilon$ as before, we can show that
\EQ{
{\cal W}_\text{inst}^ {\text{(I)}}(m_l-n_l\epsilon)
={\cal W}^\text{(II)}(\lambda_{ls})-{\cal W}^\text{(II)}(\lambda_{ls}^{(0)})\ ,
}
where we have identified
\EQ{
{\cal W}^\text{(II)}(\lambda_{ls})\equiv Y(\lambda_{ls}) = &
\log q\sum_{(ls)=1}^N\lambda_{ls}+\sum_{(ls)=1}^N\sum_{n=1}^L\big(f(\lambda_{ls}-\tilde
M_n)-f(\lambda_{ls}-M_n)\big)\\ &
+\frac12\sum_{(ls)\neq
(mp)=1}^N\big(f(\lambda_{ls}-\lambda_{mp}-\epsilon)-f(\lambda_{ls}-\lambda_{mp}
+\epsilon)\big)\ \label{DefWx2}
}
as the Yang-Yang functional for the spin chain \cite{YY}. Note that the
equations-of-motion of the functional $Y(\lambda_j)$ are the BAE
\eqref{Spinchain1}. Since the instanton contribution to
${\cal W}^\text{(I)}_\text{inst}$ at the root of baryonic Higgs branch identically vanishes
\EQ{
{\cal W}^ {\text{(I)}}_\text{inst}(m_l-\epsilon)=0\ ,
}
the complete matching of the two theories \eqref{ytt}
\EQ{
{\cal W}^ {\text{(I)}}(m_l-n_l\epsilon)-{\cal W}^ {\text{(I)}}(m_l-\epsilon)
={\cal W}^ {\text{(II)}}(\lambda_{ls})\equiv Y(\lambda_{ls})
}
requires perturbative contributions to satisfy a following
relation
\EQ{
{\cal W}_\text{pert}^ {\text{(I)}}(m_l-n_l\epsilon)-{\cal W}_\text{pert}^
{\text{(I)}}(m_l-\epsilon)
={\cal W}^ {\text{(II)}}(\lambda_{ls}^{(0)})\equiv Y(\lambda_{ls}^{(0)})\ .
\label{ftf}
}
It is rather trivial to see the matching of
the classical parts
\EQ{
{\cal W}_\text{cl}^ {\text{(I)}}(m_l-n_l\epsilon)-{\cal W}_\text{cl}^
{\text{(I)}}(m_l-\epsilon)
=\log q\sum_{(ls)=1}^N\lambda_{ls}^{(0)}\ ,
}
where
\EQ{
{\cal W}_\text{cl}^ {\text{(I)}}(a_l)=-\frac{\log
q}{2\epsilon}\sum_{l=1}^La_l^2\ .
}
%
The one-loop contribution is given by
\EQ{
{\cal W}_{\text{1-loop}}^
{\text{(I)}}(a_l)=\sum_{l,n}\big[\omega_\epsilon(a_l-\tilde
m_n-\epsilon)+\omega_\epsilon(a_l-m_n)-\omega_\epsilon(a_l-a_n)\big]\ ,
}
where $\omega_\ep(x)$ satisfies
$\frac{d\omega_{\ep}(x)}{dx}=-\log\Gamma(1+x/\ep)$.
It needs much elaboration, discussed in details in \cite{DHL},
to show that
\EQ{
{\cal W}_\text{1-loop}^{\text{(I)}}(m_l-n_l\epsilon)-
{\cal W}_\text{1-loop}^{\text{(I)}}(m_l-\epsilon)
={\cal W}^\text{(II)}\big(\lambda^{(0)}_j\big)-\log q\sum_{(ls)}\lambda^{(0)}_{ls}\ ,
}
which completes the proof of the conjectured duality in \cite{DHL} between
in Theories I and II (\ref{ytt}).

Let us finish this section by commenting on the VEVs of the chiral operators
$\hat{\cO}_k=\Tr \varphi^k$. It was was proposed in \cite{DHL} that
these are related in a simple way
to the conserved charges of the associated spin chain which correspond
to the coefficients of the polynomial $t(\lambda)$ appearing in the
Baxter eqn above. This is a natural generalisation of the usual
relation between the corresponding VEVs and the coefficients in the polynomial
$T(x)$ appearing
in the Seiberg-Witten curve (\ref{swc}) of the undeformed
$\epsilon=0$ case. The proposal of
\cite{DHL} can be explicitly recast as,
\EQ{
\big\langle \Tr  \varphi^k\big\rangle_\text{DHL}&=
\int_{\cC}\frac{d\lambda}{2\pi
i} \lambda^k
\frac{d}{d\lambda}\log\left(\frac{Q(\lambda+\epsilon)}{Q(\lambda)}\frac{Q_0(\lambda)}{Q_0(\lambda+\ep)}\right)\\
&\qquad\qquad+\int_{\cC}\frac{d\lambda}{2\pi i} \lambda^k
\frac{d}{d\lambda}\log
\left(1+q\frac{a(\lambda)Q(\lambda-\epsilon)}{d(\lambda)Q(\lambda+\ep)}\right)\
.
\label{vevSW}
}
On the other hand
we can calculate the expectation values directly using the instanton
calculus instead. Indeed
the Nekrasov partition function with operators $\hat{\cO}_k$ inserted can be
evaluated readily using the saddle point approach described above
\cite{Italians1}. In our notation this yields:
\footnote{Here we have removed the perturbative
pieces, and taken into account the mapping between the parameters in Theory I
and II as given in (\ref{Id1}).}
\begin{equation}
\big\langle \Tr \varphi^k\big\rangle_\text{SC}=\int_{\cC}\frac{d\lambda}{2\pi
i}\, \lambda^k
\frac{d}{d\lambda}\log\left(\frac{Q(\lambda+\epsilon)}{Q(\lambda)}\frac{Q_0(\lambda)}{Q_0(\lambda+\ep)}\right)\,.\label{vevSC}
\end{equation}
Which reproduces the first term of (\ref{vevSW}) but not the second.
Here the contour $\cC$ encloses the entire complex plane, hence all the zeros
in $Q(\lambda)$, $Q(\lambda+\ep)$, $Q_0(\lambda)$ and $Q_0(\lambda+\ep)$,
$Q_0(\lambda)$ is defined as $Q(\lambda)$, with $\lambda_i\to \lambda_i^{(0)}$.

For the case of ${\cal N}=2$ SQCD with gauge
group $SU(L)$ and $N_{F}<L$ fundamental flavours
the two corresponding
definitions were shown to be equivalent in \cite{Italians1}
(see in particular Eqn (46) in this reference). However, the
equivalence does not hold for $N_{F}\geq L$ and in particular does not
hold in the present case $N_{F}=2L$. This reflects a well known
ambiguity in parametrising the Coulomb branch first uncovered in
\cite{DKM}. Even in the undeformed case $\epsilon=0$, it is known that
the VEVs extracted from the Seiberg-Witten curve are not equal to
those obtained from direct semiclassical calculations but are related
to the latter by holomorphic operator mixings which are allowed by the
symmetries of the theory. In the present case, the VEVs conjectured in
\cite{DHL} are related by similar holomorphic mixings to those of the
direct calculation. The explicit form of these mixings can be deduced
from the equations given above but we will not consider these further here.

\section{Generalisation to Linear Quiver Theories}

\begin{figure}
\centering
\includegraphics[width=135mm]{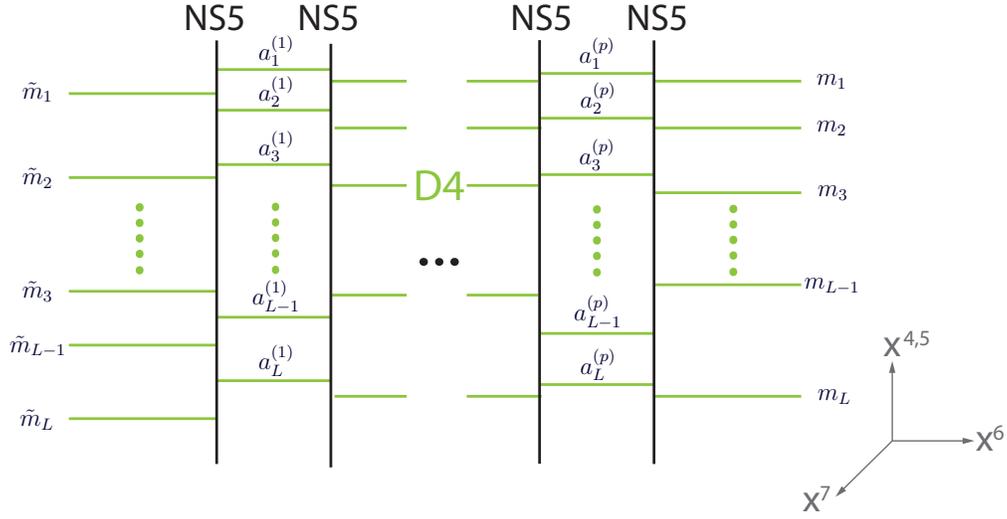}
\caption{The IIA-brane construction for Theory I in the linear quiver
case.}
\label{THEORYI}
\end{figure}
\begin{figure}
\centering
\includegraphics[width=135mm]{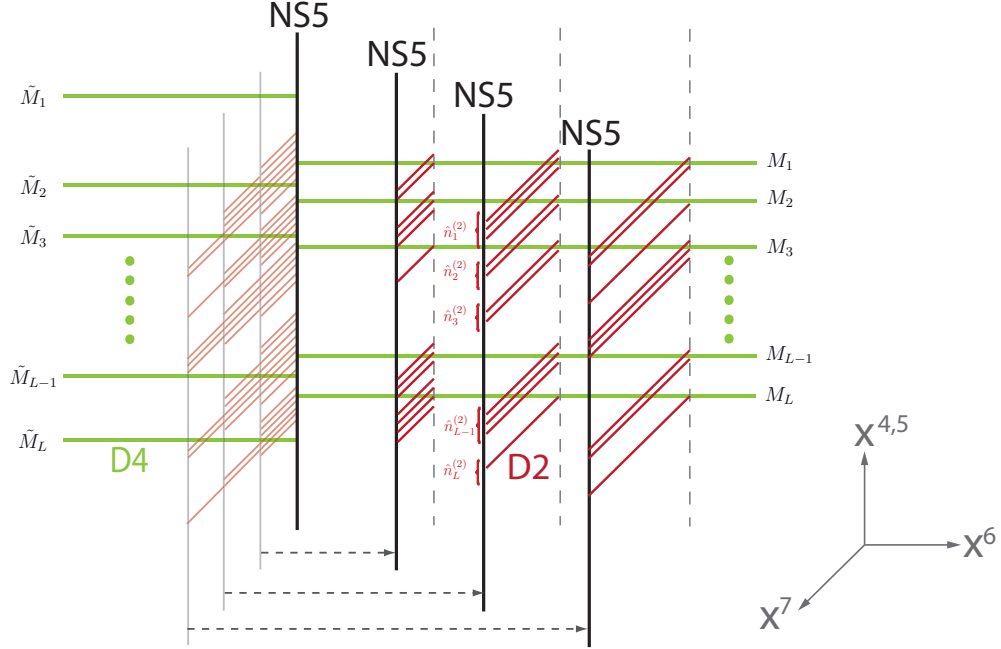}
\caption{The IIA-brane brane construction for Theory II in the linear quiver
case.}

\label{THEORYII}
\end{figure}

We can now apply these ideas to the quiver gauge theories and derive the
equations which can be interpreted as the BAE of an associated spin
system. Brane constructions of dual four- and two-dimensional theories
are shown in Figures (\ref{THEORYI}) and (\ref{THEORYII}) respectively.
As Theory I, we
will consider the $A_{p}$ linear quiver theory in four dimensions
with gauge group $SU(L)^p$ and
bi-fundamental hypermultiplets between the nodes of mass $\mu_I$,
$I=1,\ldots,p-1$ and the first and last node have $L$(anti-) fundamental
hypermultiplets of mass $-\tilde m_l$ and $-m_l+\epsilon$, respectively.
The contribution from a bi-fundamental hypermultiplet of mass $\mu_I$ charged
under the $I^\text{th}$ and $I+1^\text{th}$ $SU(L)$ factors of the gauge group
to the instanton partition function is
\EQ{
&{\cal Z}_\text{bi-fund}(\vec Y)\\ &=
\prod_{li,nj}\frac{
\Gamma\big(\epsilon_2^{-1}(x^{(I)}_{li}-x^{(I+1)}_{nj}+\mu_I)\big)}
{\Gamma\big(\epsilon_2^{-1}(x^{(0,I)}_{li}-x^{(0,I+1)}_{nj}+\mu_I)\big)}\cdot\frac{
\Gamma\big(\epsilon_2^{-1}(x^{(I+1)}_{li}-x^{(I)}_{nj}+\epsilon_1+\epsilon_2-\mu_I)\big)}
{\Gamma\big(\epsilon_2^{-1}(x^{(0,I+1)}_{li}-x^{(0,I)}_{nj}+\epsilon_1+\epsilon_2-\mu_I)\big)}
\ .
\label{vmm2}
}
There is a subtlety here, explained in \cite{AGT}, that the contribution is
not symmetric under interachanging $I$ and $I+1$, rather one must also change
$\mu_I\to\epsilon_1+\epsilon_2-\mu_I$.
Taking the NS limit as before gives rise to the following terms in the
instanton action
${\cal Y}_{I,I+1}(x_j)-{\cal Y}_{I,I+1}(x_j^{(0)})$ where
\EQ{
{\cal
Y}_{I,I+1}(x_j)=\sum_{li,nj}\Big(f\big(x_{li}^{(I)}-x_{nj}^{(I+1)}+\mu_I\big)+f\big(x_{li}^{(I+1)}-x_{nj}^{(I)}+\epsilon-\mu_I\big)\Big)\
.
}

As previously, the instanton action functional can be written in terms of a set
of instanton densities $\rho_I(x)$, $I=1,\ldots,p$, which are constant between
the points $[x_{li}^{(I)},x_{li}^{(0,I)}]$ as
\EQ{
{\cal H}_{\rm inst}[\rho_I]=-\frac12\int dx\,dy\,\rho_I(x)\mathfrak G_{IJ}(x-y)\rho_J(y)
+\int dx\,\rho_I(x)\log\big(q_I{\mathfrak R}_I(x)\big)\ ,
}
where the non-vanishing components of the kernel are
\EQ{
\mathfrak G_{II}(x)=\frac{d}{dx}\log\Big(\frac{x-\epsilon}{x+\epsilon}\Big)\
,\qquad
\mathfrak G_{I,I+1}(x)=\mathfrak
G_{I+1,I}(-x)=\frac{d}{dx}\log\Big(\frac{x+\mu_I}{x-\epsilon+\mu_I}\Big)\ .
}
We also define
\EQ{\label{DefRIx}
&\mathfrak
R_I(x)=\frac{P_{I-1}(x+\epsilon-\mu_{I-1})P_{I+1}(x+\mu_I)}{P_I(x)P_I(x+\epsilon)}\
,\qquad 1<I<p\ ,\\
&\mathfrak R_1(x)=\frac{A(x)P_2(x+\mu_1)}{P_1(x)P_1(x+\epsilon)}\ ,\qquad
\mathfrak
R_p(x)=\frac{P_{p-1}(x+\epsilon-\mu_{p-1})D(x+\epsilon)}{P_p(x)P_p(x+\epsilon)}\
,
}
where $P_I(x)=\prod_{l=1}^L(x-a^{(I)}_l)$.
The saddle-point equations are simple to write down.
When the quantisation conditions are imposed
\EQ{
a_l^{(I)}=m_l-n^{(I)}_l\epsilon-\sum_{J=I}^{p}\mu_J\ ,
\qquad \mu_p = 0 \ ,
\label{quant2}
}
one can again show the degeneration of intervals
\EQ{
x_{li}^{(I)}=x_{li}^{(0,I)}=a^{(I)}_l+(i-1)\epsilon\ ,\qquad i\geq n^{(I)}_l\ ,
}
leading to truncation of the saddle-point equations.
The root of baryonic Higgs branch in this linear quiver case
is now located at
\EQ{a_l^{(I)}=m_l-\ep-\sum_{J=I}^{p}\mu_J\ ,
\label{BHRoot}}
where ${\mathfrak R}_I(x)$ in (\ref{DefRIx}), or equivalently
instanton partition function vanish identically,
due to the additional zero modes that pop up the Higgs branch moduli.

Defining again the truncated quantities
\EQ{
\hat {\mathfrak Q}_I(x)=\prod_{l=1}^L \prod_{i=1}^{n_l^{(I)}-1} (x-x^{(I)}_{li})\ ,
}
the saddle point equations become%
%
%
\begin{gather}
 -q_1\frac{\hat{\mathfrak Q}_1(x_{li}^{(1)}-\epsilon)}
 {\hat{\mathfrak Q}_1(x_{li}^{(1)}+\epsilon)}
 \frac{\hat{\mathfrak Q}_{2}(x_{li}^{(1)}+\mu_{1})}
 {\hat{\mathfrak Q}_2(x_{li}^{(1)}-\epsilon+\mu_{1})}
 =\frac{D(x_{li}^{(1)}+\sum_{J=1}^{p-1}\mu_J+2\epsilon)}{A(x_{li}^{(1)})}\ ,
 \nonumber \\
 -q_I\frac{\hat{\mathfrak Q}_{I-1}(x_{li}^{(I)}+\epsilon-\mu_{I-1})}
 {\hat{\mathfrak Q}_{I-1}(x_{li}^{(I)}-\mu_{I-1})}
 \frac{\hat{\mathfrak Q}_I(x_{li}^{(I)}-\epsilon)}
 {\hat{\mathfrak Q}_I(x_{li}^{(I)}+\epsilon)}
 \frac{\hat{\mathfrak Q}_{I+1}(x_{li}^{(I)}+\mu_{1})}
 {\hat{\mathfrak Q}_{I+1}(x_{li}^{(I)}-\epsilon+\mu_{I})}=1\ ,
 \qquad ( 1< I < p ) \nonumber \\
 -q_p\frac{\hat{\mathfrak Q}_{p-1}(x_{li}^{(p)}+\epsilon-\mu_{p-1})}
 {\hat{\mathfrak Q}_{p-1}(x_{li}^{(p)}-\mu_{p-1})}
 \frac{\hat{\mathfrak Q}_p(x_{li}^{(p)}-\epsilon)}
 {\hat{\mathfrak Q}_p(x_{li}^{(p)}+\epsilon)}=1\ .
 \label{dfg}
\end{gather}
With the dictionary below
\EQ{
&x^{(I)}=\lambda^{(I)}-\sum_{J=I}^{p-1}\big(\mu_J-\frac12\epsilon)-\frac12\epsilon\
,\\
&
M_l=m_l-\frac{p+2}2\epsilon\ ,\qquad\tilde M_l=\tilde
m_l+\sum_{J=1}^{p-1}\big(\mu_J-\frac12\epsilon)+\frac12\epsilon\ ,
}
the above equations (\ref{dfg}) are exactly the BAE of
an $SL(p+1,{\mathbb R})$ spin chain
\EQ{
-q_I\prod_{J=1}^p\frac{{{Q}}_J(\lambda_j^{(I)}-\tfrac12\epsilon
C_{IJ})}{{Q}_J(\lambda_j^{(I)}+\tfrac12\epsilon C_{IJ})}=\begin{cases}
\frac{d(\lambda_j^{(1)})}{a(\lambda_j^{(1)})}  & I=1\\ 1 & I>1\ ,\end{cases}
\label{bae2}
}
where $a(\lambda)$ and $d(\lambda)$ are defined in \eqref{sqq} and $C_{IJ}$ is
the Cartan matrix of the Lie algebra associated to $SL(p+1)$,
\EQ{
C_{IJ}=2\delta_{IJ}-\delta_{I,J+1}-\delta_{I,J-1}\ .
}
The classical instanton string solutions,
\EQ{
x_{li}^{(I,0)}=a^{(I)}_l+(i-1)\epsilon=m_l-n^{(I)}_l\epsilon-\sum_{J=I}^{p-1}\mu_J+(i-1)\epsilon\
,
}
$i=1,\ldots,n_l^{(I)}-1$,
are related to classical Bethe roots
\EQ{
\lambda_{(ls)}^{(I,0)}=M_l-(s-1)\epsilon+\frac{I-1}2\epsilon\ ,
}
by $s=n_l^{(I)}-i$.
It is straightforward to show that the instanton action matches the Yang-Yang
functional of the spin chain generalizing \eqref{onshellH} in an obvious way.

Theory II in correspondence is therefore
a two dimensional ${\mathcal N}=(2,2)$ super QCD with quiver gauge group
$\prod_{I=1}^p U(N_I)$ with $N_I = \sum_{l=1}^L (n_l^{(I)} -1)$.
Theory II has the matter content of
one adjoint hypermultiplet with twisted mass $\epsilon$ for each $U(N_I)$,
bi-fundamental of twisted mass $\frac12\epsilon$
under $U(N_I)\times U(N_{I+1})$,
and $L$ fundamental hypermultiplet of masses $M_l$ and anti-fundamental $L$
anti-fundamental of masses $\tilde M_l$ under $U(N_1)$.
As depicted in Figure (\ref{THEORYII}),
one can show $n_l^{(I)} -1  = \sum_{J=I}^p \hat{n}_l^{(J)} $ or
equivalently $N_I = \sum_{J=I}^p \sum_{l=1}^L \hat
n_l^{(J)}$, where $\hat n_l^{(J)}$ denotes
a number of D2-branes stretched between $l^\text{th}$ D4-brane and
$J^\text{th}$ NS5-brane. This relation is compatible with an
interpretation of the duality in terms of
the refined geometric transition proposed in \cite{DHL}.

The twisted superpotential/Yang-Yang functional for Theory II
can be quite straightforwardly
written down, using the results in \cite{OrlRef}, and the BAE arising from the
F-term on-shell condition is precisely \eqref{bae2} above with
\EQ{
\hat q_I=(-1)^{N_I+1} q_I\ .
}

In order to complete the duality between the two sides, we need to show that
the perturbative pieces match; that is,
\EQ{
{\cal
W}_\text{pert}^{\text{(I)}}\Big(m_l-n_l^{(I)}\epsilon-\sum_{J=I}^{p-1}\mu_J\Big)-
{\cal
W}_\text{pert}^{\text{(I)}}\Big(m_l-\epsilon-\sum_{J=I}^{p-1}\mu_J\Big)={\cal
W}^ {\text{(II)}}\big(\lambda_j^{(I,0)}\big)\ ,
}
which generalizes \eqref{ftf}.
The matching of the classical  contributions is guaranteed by the identity
\EQ{
\Big(m_l-n_l^{(I)}\epsilon-\sum_{J=I}^{p-1}\mu_J\Big)^2-\Big(m_l-\epsilon-\sum_{J=I}^{p-1}\mu_J\Big)^2
=-2\epsilon\sum_{s=1}^{n^{(I)}_l-1}\lambda^{(I,0)}_{(ls)}\ .
}
The one-loop contribution is given by
\EQ{
{\cal W}_{\text{1-loop}}^
{\text{(I)}}(a^{(I)}_l)&=\sum_{ln}\big[\omega_\epsilon(a_l^{(1)}-\tilde
m_n)+\omega_\epsilon(a_l^{(p)}-m_n-\epsilon)+
\omega_\epsilon(a_l^{(I)}-a_n^{(I)})
\big]\\ &+\sum_{I=1}^{p-1}\sum_{ln}\omega_\epsilon(a^{(I)}_l-a^{(I+1)}_n+\mu_I)
}
and after some tedious work, one can show that
\EQ{
&{\cal
W}_\text{1-loop}^{\text{(I)}}\Big(m_l-n_l^{(I)}\epsilon-\sum_{J=I}^{p-1}\mu_J\Big)-
{\cal W}_\text{1-loop}^{\text{(I)}}\Big(m_l-\epsilon-\sum_{J=I}^{p-1}\mu_J\Big)
\\ &\qquad\qquad={\cal W}^\text{(II)}\big(\lambda^{(I,0)}_j\big)-\log
q\sum_{Ij}\lambda^{(I,0)}_j\ ,
}
as required,
and this completes the proof of the duality for the finite quiver theories.

\vskip 2cm
\noindent {\bf\large Acknowledgement}
\vskip 2mm
HYC is generously supported in part by NSF CAREER Award No. PHY-0348093, DOE
grant
DE-FG-02-95ER40896, a Research Innovation Award and a Cottrell Scholar Award
from Research Corporation, and a Vilas Associate Award from the University of
Wisconsin.

\end{document}